\documentclass{elsart2}

\usepackage{amsmath}
\usepackage{amssymb}
\usepackage{amsfonts}
\usepackage{epsfig,graphicx}
\usepackage{subfigure}
\usepackage{graphicx}
\usepackage{rotating}
\usepackage{hhline}
\usepackage{bm}
\usepackage{color}
\usepackage[sort&compress,square]{natbib}

\def\vplus{|{\rm vac}\rangle_+}
\def\vzero{|{\rm vac}\rangle_0}
\def\sst{\scriptscriptstyle}
\def\hf{{\textstyle{1\over2}}}
\newcommand{\Nc}{N_{\textrm{c}}}
\newcommand{\Nf}{N_{\textrm{f}}}
\newcommand{\Tr}{{\textrm Tr}}

\begin{document}

\begin{frontmatter}



\title{\vspace{-0.5cm}\hfill{\scriptsize IPPP/08/80}\\[-0.4cm]\hfill{\scriptsize DCPT/08/160}\\[0.8cm]Metastable SUSY Breaking\\[-0.1cm] -- \\[-0.1cm]Predicting the Fate of the Universe}

\vspace{0.3cm}
\author{Joerg Jaeckel}

\address{Institute for Particle Physics Phenomenology, \\Department of Physics, Durham University, Durham DH1 3LE, UK}

\begin{abstract}
The possibility that supersymmetry (SUSY) could be broken in a
metastable vacuum has recently attracted renewed interest. In these
proceedings we will argue that metastability is an attractive and
testable scenario. The recent developments were triggered by the
presentation of a simple and calculable model of metastable SUSY
breaking by Intriligator, Seiberg and Shih (ISS), which we will
briefly review. One of the main questions raised by metastability
is, why did the universe end up in this vacuum. Using the ISS model
as an example we will argue that in a large class of models the
universe is automatically driven into the metastable state during
the early hot phase and gets trapped there. This makes metastability
a natural option from the cosmological point of view. However, it
may be more than that. The phenomenologically required gaugino
masses require the breaking of R-symmetry. However, in scenarios
with a low supersymmetry breaking scale, e.g., gauge mediation a
powerful theorem due to Nelson and Seiberg places this at odds with
supersymmetry breaking in a truely stable state and metstability
becomes (nearly) inevitable. Turning around one can now
experimentally test whether gauge mediation is realised in nature
thereby automatically testing the possibility of a metastability of
the vacuum. Indeed, already the LHC may give us crucial information
about the stability of the vacuum.
\end{abstract}

\end{frontmatter}

% main text
\section{Introduction}
\label{}
Supersymmetry is still one of the prime candidates for physics beyond the standard model. It has many desirable consequences such as the
taming of the quadratic divergences that lead to the hierarchy problem for the Higgs mass, or leading to a better
unification of the three gauge couplings. However, it has one small defect. Nature does not appear to be supersymmetric.
For example, there is no massless photino that would be the supersymmetric partner of the photon.
Accordingly supersymmetry must be a broken symmetry.

It turns out, however, breaking supersymmetry is not easy. A number of powerful theorems severely restrict the options for breaking supersymmetry in
a stable vacuum.
One of them is the Nelson-Seiberg theorem \cite{Nelson:1993nf}. As we will discuss in detail in Sect.~\ref{inevitable} this theorem tells us
that if supersymmetry is broken in a true ground state we have a so-called R-symmetry that again forbids gaugino masses\footnote{Strictly
speaking a U(1) R-symmetry forbids Majorana masses for the gauginos which have non-vanishing R-charge. For a recent attempt to construct
a model with Dirac gaugino masses see~\cite{Amigo:2008rc}.}, or if the R-symmetry is spontaneously broken we have a massless Goldstone boson, the R-axion, which is
also ruled out by observation.

Breaking supersymmetry in a metstable state\footnote{For a precursor see~\cite{Ellis:1982vi}.} allows us to circumvent these constraints. In these notes we will explore this possibility and
address some of the questions that immediately come to mind. Are there simple models realizing metastable supersymmetry breaking (Sect.~\ref{ISS})?
Does the metastable state live long enough (Sect.~\ref{ISS})? How did the universe end up in the metatstable vacuum (Sect.~\ref{thermal})?
But in Sect.~\ref{inevitable} we will also turn around and ask: Can we find out whether we live in a metstable vacuum without actually making the transition to
the stable vacuum (and in consequence facing the end of the world as we know it)? At least in some cases the, perhaps surprising, answer to this question will
be, yes. Indeed already the LHC may allow us to shed light on the question of metastability.

\section{The ISS model}\label{ISS}

%\subsection{Set-up of the model}
The model proposed by Intriligator, Seiberg and Shih~\cite{ISS} is an ${\mathcal N}=1$ Super-QCD with $\Nc$ colors and $\Nf$ massive flavors
$Q^{a}_{i}$ and $\tilde{Q}^{i}_{a}$ ($i,j,\ldots$ denote color and $a,b,\ldots$ flavors).
The tree-level superpotential is simply the mass term
\begin{equation}
W^{\rm tree}_{\rm micro}=m_{ab} \tilde{Q}^{i}_{a}Q^{b}_{i}.
\end{equation}
In principle the masses of different flavors could be different, but for simplicity we will consider an $SU(\Nf)$ flavor symmetry and accordingly a mass term
\begin{equation}
m_{ab}=m{\mathbf{1}}_{ab}.
\end{equation}
For
\begin{equation}
b^{\rm micro}_{0}=3\Nc-\Nf>0
\end{equation}
this theory is asymptotically free and strongly coupled in the infrared. As in ordinary QCD this makes it difficult to study this theory directly in the low
energy regime.

Fortunately, in the range $\Nc+1\leq\Nf<\frac{3}{2}\Nc$ there exists an infrared free Seiberg dual~\cite{Seiberg1,Seiberg2,IS} of this theory that describes physics in the infrared.
This {\emph{macroscopic}} Seiberg dual is, similar to a pion model for QCD, a description of low energy effects in more suitable degrees of freedom.
The macroscopic Seiberg dual of our microscopic $SU(\Nc)$ gauge theory\footnote{The microscopic and macroscopic Seiberg dual are
also often called the electric and magnetic Seiberg dual, respectively.} with $\Nf$ flavors is again a
$SU(N)$ gauge theory coupled to $\Nf$ flavours of chiral superfields $\varphi^c_i$ and $\tilde{\varphi}^i_c$ transforming in the
fundamental and the anti-fundamental representations of the gauge group; $c=1,\ldots,N$ and $i=1,\ldots, \Nf.$
However $N$ is now given by
\begin{equation}
N=\Nf-\Nc.
\end{equation}
There is also an
$\Nf \times \Nf$ chiral superfield $\Phi^i_j$ which is a gauge singlet.
The number of flavours is taken to be large, $\Nf  > 3N,$ such that the $\beta$-function for the gauge coupling is positive,
\begin{equation}
b_0 = 3N - \Nf < 0.
\label{bzero}
\end{equation}
The theory is free in the IR and strongly coupled in the UV where it develops a Landau pole at the energy-scale $\Lambda_L$.
The condition \eqref{bzero} ensures that the theory is weakly coupled at scales $E \ll \Lambda_L$,
thus its low-energy dynamics as well as the vacuum structure is under control.
In particular, this guarantees a robust understanding of the theory
in the metastable SUSY breaking vacuum found in \cite{ISS}. This is one of the key features of the ISS model.
We will now continue all of our discussion in the low energy macroscopic description.

The tree-level superpotential of the ISS model is given by
\begin{equation}
W_{\rm cl}\, =\, h\, \Tr_{\sst \Nf} \varphi \Phi \tilde{\varphi}\, -\, h\mu^2\, \Tr_{\sst \Nf} \Phi
\label{Wcl}
\end{equation}
where $h$ and $\mu\sim\sqrt{m\Lambda_{L}}$ are constants. The usual holomorphicity arguments imply that the superpotential
\eqref{Wcl} receives no corrections in perturbation theory. However, there is a non-perturbative contribution
to the full superpotential of the theory, $W=W_{\rm cl} + W_{\rm dyn},$ which
is generated dynamically. $W_{\rm dyn}$ was determined in \cite{ISS} and is given by
\begin{equation}
W_{\rm dyn}\, =\, N\left( h^\Nf \frac{\det_{\sst \Nf} \Phi}{\Lambda_{L}^{\Nf-3N}}\right)^\frac{1}{N}.
\label{Wdyn}
\end{equation}
This dynamical superpotential is exact, its form is uniquely determined by the symmetries of the theory
and it is generated by instanton-like configurations.

The authors of \cite{ISS} have studied the vacuum structure of the theory and established the
existence of the metastable vacuum $\vplus$ characterised by
\begin{equation}
\langle \varphi \rangle =\, \langle \tilde{\varphi}^T \rangle = \, \mu \left(
{\mathbf{1}}_{N}, 0_{\Nf-N}\right)^{T} \ , \quad
\langle \Phi \rangle = \, 0 \ ,
\qquad V_+ = \, (\Nf-N)|h^2 \mu^4|
\label{vac+}
\end{equation}
where $V_+$ is the classical energy density in this vacuum.
Supersymmetry is broken
since $ V_+ >0.$ and the scale of supersymmetry breaking is given by $(V_+)^{1/4}\sim\mu$.
In this vacuum the $SU(N)$ gauge group is Higgsed
by the vevs of $\varphi$ and $\tilde{\varphi}$ and the gauge degrees of freedom
are massive with $m_{\rm gauge} = g \mu.$
%This supersymmetry breaking vacuum $\vplus$ originates from the so-called
%rank condition, which implies that there are no solutions to the F-flatness
%equation $F_{\Phi^j_i}=0$ for the classical superpotential $W_{\rm cl}$ in \eqref{Wcl},
%\begin{equation}
%F_{\Phi^j_i} =\, %\frac{\partial W_{\rm cl}}{\partial \Phi^j_i} =\,
%h(\tilde{\varphi}^j_c \varphi_i^c - \mu^2 \delta^J_i) \neq \, 0.
%\label{rank}
%\end{equation}
%The non-perturbative superpotential of \eqref{Wdyn} gives negligible contributions to the
%effective potential around this vacuum and can be ignored there.
In \cite{ISS} it was argued
that the vacuum \eqref{vac+} has no tachyonic directions, is classically stable, and quantum-mechanically
is long-lived.

\begin{figure}[t]
    \begin{center}
        \includegraphics[width=6cm]{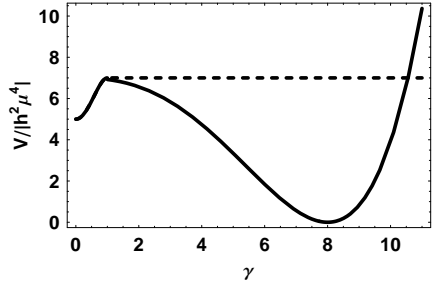}
    \end{center}
    \caption{Effective potential $V (\gamma)$ of Eq.~\eqref{VzeroT} as a function of
    $\gamma = \Phi/\mu$. For the SUSY preserving vacuum $\vzero$ we chose $\gamma=\gamma_0=8$.
    %The SUSY breaking metastable minimum $\vplus$ is always at $\gamma=0,$ and the top of the barrier is
    %always at $\gamma=1.$
    The dashed line shows the potential if we neglect the non-perturbative contribution $W_{dyn}$ to the superpotential.
    We have taken the minimal allowed values for $N$ and $\Nf$, $N=2$, $\Nf=7$. }
    \label{fig:recurs}
\end{figure}

In addition to the metastable SUSY breaking vacuum $\vplus$,
there exists a SUSY preserving stable vacuum\footnote{In fact there are precisely $\Nf-N=\Nc$ of such vacua
differing by a phase $e^{2\pi i/(\Nf-N)}$ as required by the Witten index~\cite{Witten:1982df} of the microscopic Seiberg dual formulation.}
$\vzero$,
\begin{equation}
\langle \varphi \rangle =\, \langle \tilde{\varphi}^T \rangle = \, 0 \ , \quad
\langle \Phi \rangle = \, \Phi_0=\, \mu \gamma_0 \, {\mathbf{1}}_{\Nf} \ , \qquad \qquad V_0 = \, 0
\label{vac0}
\end{equation}
where $V_0$ is the energy density in this vacuum and
\begin{equation}
\gamma_0 =\, h \epsilon^{-\frac{\Nf-3N}{\Nf-N}} \ , \quad {\rm and} \quad
\epsilon :=\, \mu/\Lambda_L \ll \, 1.
\label{gamma0}
\end{equation}
This vacuum was determined in \cite{ISS} by solving the F-flatness
conditions for the complete superpotential $W=W_{\rm cl}+W_{\rm dyn}$ of the theory.
In the vicinity of $\vzero$ the non-perturbative superpotential of \eqref{Wdyn}
is essential and allows to solve
$F_{\Phi^j_i}=0$ equations. Thus, the appearance of the SUSY preserving vacuum
\eqref{vac0} can be interpreted in our macroscopic dual description as a non-perturbative
or dynamical restoration of supersymmetry \cite{ISS}. This is also shown in Fig.~\ref{fig:recurs}.

Indeed the existence of $\Nc$ supersymmetric vacua in super-QCD with massive vector-like flavors is guaranteed
by the non-vanishing Witten index~\cite{Witten:1982df}. Allowing ourselves to live in a metastable vacuum makes a simple super-QCD like the one
discussed above a viable candidate for a SUSY breaking sector. Thereby we have effectively circumvented the Witten index constraint.

It is helpful to have a simple picture of the potential as a function of the meson field $\Phi$. Minimizing the potential in all other directions and
plugging the solution into the potential we find the following (approximate) potential,
\begin{eqnarray}
  V(\gamma)\,
& = & |h^2 \mu^4|\left\{ \begin{array}{cc}
\Nf-N + \,2 N \gamma^2(1- \hf\gamma^2)  & \qquad 0 \le \gamma\le 1 \\[0.1cm]
 \Nf\biggl( \bigl(\frac{\gamma}{\gamma_0}\bigr)^{\frac{\Nf-N}{N}}-1\biggr)^2
& \qquad 1\leq\gamma . \\
\end{array}\right.
\label{VzeroT}
\end{eqnarray}
This is plotted in Figure~\ref{fig:recurs}.
The key features of this effective potential are (1) the large distance between the two vacua,
$\gamma_0 \gg 1,$
and (2) the slow rise of the potential to the left of the SUSY preserving vacuum.
(For esthetic reasons $\gamma_0$ in Figure~\ref{fig:recurs} is actually chosen to be rather small, $\gamma=\gamma_0=8.$)

The tunnelling rate from the metastable $\vplus$
to the supersymmetric vacuum $\vzero$ can be estimated~\cite{ISS} by approximating the potential in Figure~\ref{fig:recurs} in terms of a triangle.
The action of the bounce solution in the triangular potential is of the form~\cite{LW},
\begin{equation}
S^{4D}_{\rm bounce} =\, \frac{2 \pi^2}{3 h^2}\frac{N^3}{\Nf^2} \, \left(\frac{\Phi_0}{\mu}\right)^4
%=\, \frac{2 \pi^2}{3 h^2}\frac{N^3}{\Nf^2} \, \gamma^{4}_{0}
.
%\sim \,
%\frac{h^{-6}}{\epsilon^{4(\Nf-3N)/(\Nf-N)}}
%\, \gg 1
%\qquad {\rm for} \quad \epsilon \ll 1
\end{equation}
On closer inspection the constraints imposed by this condition are
in any case very weak. One can estimate~\cite{thermal,nonthermal} that in order to guarantee a sufficient lifetime for the universe one needs
$S^{4D}_{\rm bounce} \gtrsim 400$ translating into
an extremely weak lower bound on $\Phi_{0}$,
\begin{equation}
\gamma_{0}=\left(\Phi_{0}/\mu\right)\, \gtrsim \, 3 \sqrt{h}\,
\left(\Nf^2/N^3\right)^{1/4}\lesssim 5,
\label{vacstab}
\end{equation}
where the right hand side holds for $\Nf=7$, $N=2$ and $h\leq 1$.

In the following section we will now explain why
the universe has ended up in the metastable non-supersymmetric vacuum
in the first place.

\section{Why did the universe end up in the metastable vacuum?}\label{thermal}
A crucial steps in making metastability a viable option is to establish why the universe ended up in the
metastable vacuum and not in the true vacuum which has lower energy.
The quick answer to this question is that the early universe was very hot and that in a large class of models what will later become the metastable
vacuum is preferred by entropy. Therefore, the early universe is automatically driven towards the metastable vacuum~\cite{Abel:2006cr}.
When it then cooled down it became trapped in the metastable \mbox{state~\cite{Abel:2006cr,Craig:2006kx,Fischler:2006xh}}.

Let us now add a little bit of detail to this answer.
In general we think of situations where the metastable supersymmetry breaking occurs in a sector (let us call it the MSB sector) of the full theory
which includes a supersymmetric Standard Model (MSSM). The supersymmetry breaking of the MSB sector is then
``mediated'' by couplings, messengers etc. to the MSSM part of the model.

Thermal effects then drive MSB sector automatically towards its metastable supersymmetry breaking vacuum if the following two conditions are fulfilled:
\begin{itemize}
\item[$\bullet$]{} In the early universe the supersymmetry breaking sector is in thermal equilibrium with the fields of the Standard Model.
This requires a sufficiently strong
form of mediation, e.g., direct mediation or gauge mediation.
On the other hand pure gravitational interactions would usually be too weak to ensure thermal equilibrium.
\item[$\bullet$]{} The supersymmetric vacuum has fewer light degrees of freedom then the supersymmetry breaking metastable one.
\end{itemize}

The first condition simply assures that in the early universe not only the standard model particles were at high temperature but also all fields of the
MSB sector.
The second condition however provides for the actual dynamical reason why the universe is driven towards what will later become the metastable state.
At high temperatures the effective potential corresponds to the free energy
\begin{equation}
F=E-TS
\end{equation}
where $E$ is the energy, $T$ the temperature and $S$ the entropy. From this we can immediately conclude that at high temperature states with high entropy are preferred.
Now, massless particles are more easily produced. Therefore, at a given temperature, more of them are present than very heavy
ones (at least as long as we have vanishing chemical potential). This corresponds to higher entropy. Accordingly we expect that states with many massless degrees
of freedom are preferred at high temperatures.

\begin{figure}[t]
\begin{center}
\subfigure{
\includegraphics[width=6.0cm]{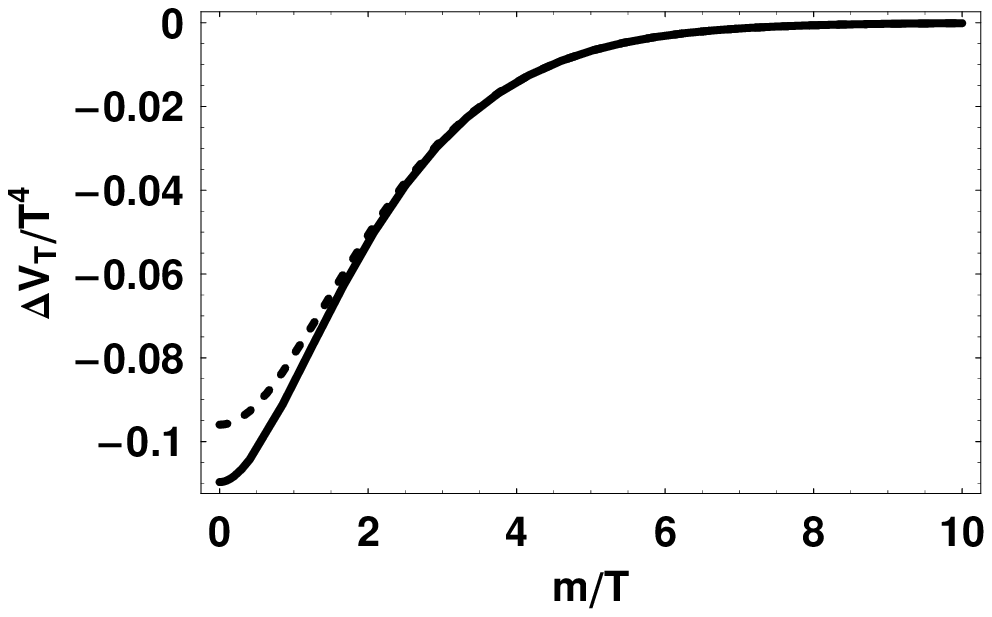}
}
\hspace{0.5cm}
\subfigure{
\includegraphics[width=6.0cm]{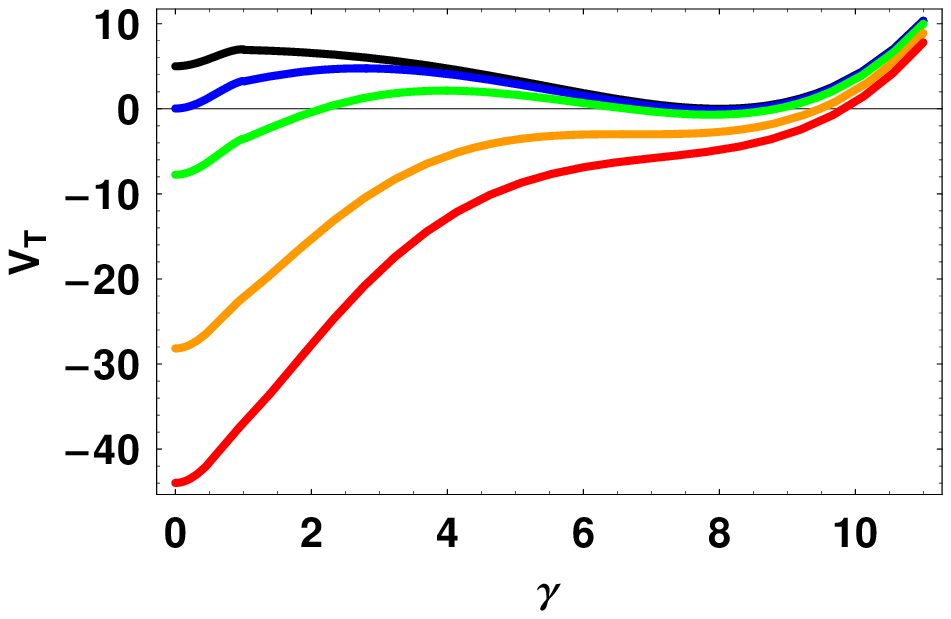}
}
\end{center}
    \caption{{\bf Left panel:} Thermal contribution $\Delta V_{T}$ to the effective potential from one species of bosons (solid line)
    and one species of fermions (dashed line) as a function of the mass $m$ of the particle species.
    {\bf Right panel:}
    Thermal effective potential for the ISS model and for different values of the temperature.
    Going from bottom to top, the red line corresponds to the temperature $T \gtrsim T_{\rm crit}$
    where we have only one vacuum at $\gamma=0$. The orange line
    corresponds to $T\approx T_{\rm crit}$ where the second vacuum appears and the classical rolling stops. The green line is in the interval
    $T_{\rm degen} < T < T_{\rm crit}$
    where one could hope to tunnel under the barrier. The blue line is at  $T \sim T_{\rm degen}$
    where the two vacua become degenerate. Finally, the black line gives the zero temperature potential where
    the non-supersymmetric vacuum at the origin becomes metastable. (We have chosen the following parameters for the ISS model $h=1$, $\gamma_{0}=8$ and
    measured everything in units of $\mu$.)}
    \label{fig:recurs2}
\end{figure}

This can be seen explicitly by calculating the 1-loop effective potential~\cite{Vthermal},
\begin{equation}
\label{dolan}
V_{T}(\Phi)=V_{T=0}(\Phi)+\frac{T^{4}}{2\pi^{2}}\sum_{i}\pm n_{i}\int^{\infty}_{0} dq \,q^{2}\ln\left(1\mp\exp\left(-\sqrt{q^{2}+m^{2}_{i}(\Phi)}\right)\right),
\end{equation}
where $n_{i}$ are the number of bosonic and fermionic species of mass $m_{i}(\Phi)$ and the upper sign is for bosons and the lower one for fermions.
At this level of approximation the field dependence of the temperature contribution (right part of Eq.~\eqref{dolan}) comes solely from the
field dependence of the masses of the particles species contributing in the sum.
In the left panel of Fig.~\ref{fig:recurs2} we plot the dependence of the temperature contribution for one species of bosons (solid line) and fermions (dashed).
Clearly at high temperature massless particles are preferred.
% \appendix

% \section{}
% \label{}

It is now clear that if our assumptions are fulfilled the metastable vacuum will be preferred at some high temperature.
Let us now briefly look at the specific situation of the ISS model described in the previous section.
In this model the metastable vacuum is located at $\Phi=0$ whereas the stable supersymmetric vacuum is at large $\Phi=\mu\gamma_{0}$.
Therefore we are looking for fields whose mass depends on $\Phi$.
The obvious candidates are the quarks $\varphi$ and $\tilde{\varphi}$.
From the superpotential \eqref{Wcl} we can read off that they obtain masses
\begin{equation}
\label{quarkmass}
m_{\varphi}=h\Phi
\end{equation}
from the Yukawa interaction with the meson field $\Phi$. Clearly the masses of the quarks are large in the supersymmetric vacuum and small in the
metastable one. Accordingly the quarks lead to a preference of the metastable vacuum at high temperatures.
In addition to the quarks we have the gauge fields of the $SU(N)$. They, too, get a mass that grows with $\Phi$. The reason for this is confinement.
For vanishing quark masses the $SU(N)$ gauge theory with $\Nf$ flavors is infrared free and does \emph{not} confine. However, for non-vanishing
quark masses, i.e. non-vanishing $\Phi$, the theory becomes a pure $SU(N)$ gauge theory below the masses of the quarks. This theory confines.
We can naively model this effect by giving the gauge particles a mass of the order of the confinement scale which we take to be the scale of the gluino condensate,
\begin{equation}
m_{\rm gauge}=\langle \lambda \lambda\rangle^{\frac{1}{3}}=h^{\frac{1}{3}}\mu\left(\frac{\gamma}{\gamma_{0}}\right)^{\frac{\Nf}{3N}}\gamma^{\frac{1}{3}}_{0}.
\end{equation}
For larger quark masses the turnaround
from the infrared free gauge theory \emph{with} quarks to the pure gauge theory happens at a higher energy scale (the quark mass).
Consequently, the confinement scale and $m_{\rm gauge}$ grow with increasing quark masses.

Adding the contributions of the quarks and the gauge bosons a plot of the total thermal effective potential is shown in the right
panel of Fig.~\ref{fig:recurs2}.
As expected at high temperatures what will later become the metastable minimum is preferred, i.e. it is the \emph{global} minimum.
Indeed at sufficiently high temperatures what will later become the supersymmetric vacuum is no minimum at all, not even a local one.
If the universe now started anywhere in field space it will automatically roll down to the field position of the
`metastable' minimum. Once there it basically gets trapped when the universe cools down.
Therefore, in these setups metastable SUSY breaking does not require any special initial conditions but the universe is automatically driven
towards this vacuum by thermal effects.

Finally, one might be worried that the temperatures required for the suppersymmetric minimum to disappear might be quite large. This turns out not to be
the case. Even if the supersymmetric minimum is extremely far away say $\gamma_{0}=10^6$ temperatures $\lesssim 10\mu$ (remember that $\mu$ is the scale
of supersymmetry breaking in the ISS sector) are sufficient.

\section{Detecting metastability}\label{inevitable}
Having shown that supersymmetry breaking is a viable option from the cosmological point of view we now want to go one
step further. We want to argue that in models of low-scale supersymmetry breaking, i.e. models where the gravitino mass
is smaller than the scale of supersymmetry breaking in the standard model sector, $m_{3/2}\ll m_{SUSY}$, is not only viable but it might
indeed be {\emph{inevitable}}~\cite{Intriligator:2007py,Abel:2008ve}.

The seeds of the inevitability of metastability in this class of models lie in an
important theorem due to Nelson and Seiberg~\cite{Nelson:1993nf}, who
identified a \emph{necessary} condition for $F$-term supersymmetry breaking: The existence of an $R$-symmetry\footnote{An
$R$-symmetry is a symmetry under which fermions and bosons of the same superfield transform differently. This arises because the superspace coordinates
$\theta$ transform non-trivially under an $R$-symmetry.}.
The problem arises because in
$R$-symmetric theories the gauginos must be massless, in conflict with experiments, which require
$m_{gaugino} \gtrsim 100$~GeV.
The dilemma is that non-vanishing gaugino masses require both supersymmetry breaking
{\em and} $R$-symmetry
breaking, but Nelson and Seiberg tell us that these two requirements are
mutually exclusive. How to get around it?

There are two logical possibilities.
One is to include in the theory a small, controlled amount of $R$-symmetry breaking.
More precisely, the Lagrangian would be of the form
\begin{equation}
\label{lagrangian}
\mathcal{L}=\mathcal{L}_{R}+\varepsilon\mathcal{L}_{R-breaking},
\end{equation}
where $\mathcal{L}_{R}$ describes a theory which preserves $R$-symmetry and breaks supersymmetry,
whereas $\mathcal{L}_{R-breaking}$ breaks $R$-symmetry, and
$\varepsilon$ is our small control parameter.
When $\varepsilon=0$, the lowest-energy state breaks supersymmetry,
and there is no supersymmetric vacuum at all,
but the gauginos are massless.
However, with a small
$\varepsilon \ne 0$, $R$-symmetry is broken explicitly.
In this case, the Nelson-Seiberg theorem requires that a
supersymmetry-preserving vacuum
appears in addition to the
supersymmetry-breaking one,
since the full theory breaks $R$-symmetry.
It is a general consequence of supersymmetry that any supersymmetric vacuum must be the state
of lowest energy. Hence, the non-supersymmetric vacuum must be
 metastable. However, it is important to note that the two vacua are separated by
a distance that goes to infinity as $\varepsilon\rightarrow 0$.
As the control parameter
$\varepsilon \to 0$, the decay rate of our false vacuum becomes
exponentially longer and longer.

The second possible way to obtain non-vanishing
gaugino masses is to break
the $R$-symmetry spontaneously.
Spontaneous (rather than explicit) breaking of $R$-symmetry does not introduce
new
supersymmetry preserving minima, and does not by itself make the supersymmetry
breaking vacuum metastable.
In particular we do not need to introduce and explain the
origins of a very small parameter $\varepsilon$, as we had
to with explicit breaking.
At the same time, gauginos acquire masses proportional to the scale of
spontaneous $R$-breaking.

Superficially then, it looks as if one might be able to avoid metastability. However, spontaneous breaking of
a global $U(1)$ $R$-symmetry leads to a Goldstone boson, the $R$-axion.
In order to avoid astrophysical and experimental bounds, the $R$-axion
must also acquire a mass, although the lower bounds on its mass are much weaker than
those on the gaugino mass: $m_{R-{\rm axion}} \gtrsim 100$~MeV, and
therefore easier to fulfill. Nevertheless, its mass means that the original
$R$-symmetry must itself be
explicitly broken by very small effects,
and according to the earlier arguments,
this again implies that the vacuum is metastable. In this case, however,
the gaugino mass is divorced from the size of the \emph{explicit} $R$-breaking parameter
$\varepsilon$, which now determines the $R$-axion mass instead.
This exhausts the logical possibilities and shows that massive gauginos
and massive $R$-axions imply metastability.

The ISS model~\cite{ISS} was an important step forward because it provides a simple, explicit and calculable way to generate a Lagrangian
of the form \eqref{lagrangian}.
In the ISS model a naturally small $R$-breaking term of the required type is generated dynamically by
quantum effects, more precisely by the anomaly of the U(1) $R$-symmetry.

This breakthrough has led to a burst of activity
building gauge-mediated models incorporating the ISS models as hidden sectors.
The complementary explicit and spontaneous approaches to model-building
were successfully incorporated with a few twists.
In the first approach, the explicit $R$-breaking of the ISS
models was not able to generate gaugino masses, so a second source
of $R$-breaking was required.
However, the smallness of this second term -
necessary for the longevity of the metastable vacuum,
turned out to be guaranteed within the ISS models if the $R$-symmetry-breaking
effects were generated at a very large energy scale,
e.g., the Planck scale~\cite{Murayama:2006yf}.

In the second approach, the gauginos are already massive and, as we discussed above,
the job of the explicit $R$-breaking is merely to give the $R$-axion a small mass
$m_{axion}\gtrsim$100 MeV. The controlled quantum effects within
all models of the ISS type are sufficient to do this, and
remarkably simple
versions of the ISS model could be found that led to the required
spontaneous $R$-breaking~\cite{Abel:2007jx},
so that gauginos
receive sufficiently large masses $m_{\rm gaugino} \gtrsim 100$~GeV.
These are explicit, credible models with metastable vacua.
The LHC will be able to produce gauginos weighing an order of
magnitude more than the present lower limit~\cite{LHC}, offering a good prospect
of testing such scenarios.

Finally, let us briefly comment on the issue of the cosmological constant: global
supersymmetry breaking always generates a large vacuum energy, much larger then the observed tiny value.
This contribution
can in principle be compensated
in supergravity
which can easily generate an additional negative contribution to the vacuum energy.  Adding this
contribution would not change our conclusions about the metastability of the vacuum.

\section{Conclusions}\label{conclusions}
In these notes we have argued that supersymmetry breaking could be realised in a vacuum that is only metastable.
Indeed simple models can be constructed where a metastable supersymmetry breaking vacuum exists and is sufficiently long-lived.
In a large class of models this is a viable scenario from the cosmological point of view, because independent of the initial state
thermal effects
{\emph{automatically}} drive the universe to what will (after the universe has cooled sufficiently) become the metastable vacuum.
In generic models of low-scale supersymmetry breaking metastability is not only viable but indeed inevitable.
If experiments find that the fundamental scale of supersymmetry breaking is low, metastability is essentially guaranteed.
The prototype example of a low scale supersymmetry breaking scenario is gauge mediation. In some cases already the LHC will be able to decide whether
gauge mediation, and in turn low scale supersymmetry breaking is realized. Therefore, the LHC may be able to give us a first glimpse on the ultimate
fate of the universe.

\section*{Acknowledgements}
The author wishes to thank the organizers of the SEWM 2008 conference for an interesting meeting.
Moreover, he is indebted to S.~Abel, C.~Durnford, J.~Ellis and V.~V.~Khoze for stimulating discussions and fruitful collaboration on the issues
discussed in these proceedings.

\end{document}